\documentstyle[preprint,aps]{revtex}
\begin{document}
\title{Direct Instantons and Nucleon Magnetic 
Moments\thanks{DOE/ER/40762-176, UMD PP\# 99-088} }
\author{M. Aw\thanks{Present address: Department of Physics, Carnegie Mellon 
University, Pittsburgh, PA 15213. E--mail: mountaga@cmu.edu} and M.K. Banerjee}
\address{Department of Physics, University of Maryland, College Park, MD
20742-4111}
\author{H. Forkel}
\address{Institut f{\"u}r Theoretische Physik, Universit{\"a}t Heidelberg, 
D-69120 Heidelberg, Germany}
\date{\today}
\maketitle
\begin{abstract}
We calculate the leading direct-instanton contributions to the 
operator product expansion of the nucleon correlator in a magnetic 
background field and set up improved QCD sum rules for the nucleon 
magnetic moments. Remarkably, the instanton contributions are found 
to affect only those sum rules which had previously been considered 
unstable. 
The new sum rules show good stability and reproduce the experimental 
values of the nucleon magnetic moments with values of $\chi$, the 
quark condensate magnetic susceptibility, consistent with other estimates.
\end{abstract}

For over two decades QCD instantons have been associated with fundamental 
aspects of strong interaction physics, as, for example, with 
the $\theta$-vacuum \cite{jac76}, the issue of strong CP violation 
\cite{jac76}, and the anomalously large $\eta'$ mass \cite{tho76}. 
Unequivocal and quantitative evidence for their role in hadron 
structure, however, turned out to be much harder to establish. This 
is mainly due to the complexity involved in dealing with interacting 
instanton ensembles and their coupling to other vacuum fields over 
large distances. 

Instanton vacuum models \cite{sch98} attack the first part of this 
problem directly, by approximating the field content of the vacuum 
as a superposition of solely instantons and anti-instantons. This  
approach has been developed for more than a decade and can describe 
an impressive amount of hadron phenomenology \cite{sch98}. More 
recently, QCD lattice simulations began to complement such vacuum 
models by isolating instantons in equilibrated lattice configurations 
and by studying their size distribution and their impact on hadron 
correlators \cite{lat}. While the results obtained with different, 
currently developed lattice techniques have not yet reached quantitative 
agreement, they do confirm the overall importance of instantons and 
some bulk properties of their distribution in the vacuum. 

Another approach towards linking the instanton component of the 
vacuum to hadron properties has been developed over the last years 
by generalizing the nonperturbative operator product expansion (OPE) 
and QCD sum rule techniques \cite{FB,for95,for97}. While its range 
of applicability is more limited than that of instanton vacuum models 
and of lattice calculations, it avoids the need for large-scale 
computer simulations and takes, in contrast to instanton models, all 
long-wavelength vacuum fields and also perturbative fluctuations into 
account. Furthermore, the approach is largely model-independent and 
allows the study of instanton effects in a fully analytic and, 
therefore, rather transparent fashion. 

In the present paper, we adapt this approach to properties which 
characterize the response of the hadronic system to a weak external 
field. Specifically, we calculate the direct-instanton contributions 
to the nucleon correlator in the presence of a constant electromagnetic 
field. With the help of background-field sum-rule techniques due to 
Ioffe and Smilga \cite{IS} and Balitsky and Yung \cite{BY}, we then 
study previously neglected instanton effects in the QCD sum rules for 
the magnetic moments of the nucleon.

As already mentioned, our work is based on the nucleon correlation 
function 
\begin{equation}
i\int d^{4}x \, e^{ipx}{\langle0|T \eta(x) \bar{\eta}(0)|0\rangle}_{F}
=\Pi_0(p)+\sqrt{4\pi\alpha}\Pi_{\mu\nu}(p)F^{\mu\nu}, \label{corr1}
\end{equation}
in the background of a constant electromagnetic field $F_{\mu\nu}$.
The interpolating fields $\eta(x)$ with proton or neutron quantum 
numbers \cite{I} are composite operators of massless up and down 
quark fields:
\begin{equation}
\eta_{p}(x)=[u^{a^{T}}(x)C\gamma_{\alpha}u^{b}(x)]\gamma_{5}
\gamma^{\alpha} d^{c}(x)\epsilon^{abc}, \qquad
\eta_{n}=\eta_{p}(u \leftrightarrow d). \label{interpol}
\end{equation}
For the application in QCD sum rules we need a theoretical description 
of the correlator (\ref{corr1}) at momenta $s=-p^2 \simeq 1 {\rm 
GeV}^2$, i.e. at distances $x \lesssim 0.2 \, {\rm fm}$.

The information on the magnetic moments is contained in the second 
term on the right-hand side of Eq. (\ref{corr1}). It characterizes 
the linear response of the nucleon to the external field and can be 
decomposed into three independent Lorentz and spinor structures:
\begin{equation}
\Pi_{\mu\nu}(p) =  ( \rlap/p\sigma_{\mu\nu} + 
\sigma_{\mu\nu}\rlap/p ) \, \Pi_1 (p^2) + i (\gamma_{\mu}p_{\nu} 
- \gamma_{\nu}p_{\mu}) \rlap/p  \, \, \Pi_2 (p^2) 
+ \sigma_{\mu\nu} \, \Pi_3 (p^2). \label{linresp}
\end{equation}
Note that the invariant amplitude $\Pi_1$ corresponds to the 
chirally-even part of the correlator, while $\Pi_{2}$ and $\Pi_{3}$ 
are associated with the chirally-odd part.  

The nonperturbative OPE \cite{SVZ,IS} of the above correlator at small 
distances can be generated by splitting each diagram contributing  
to (\ref{linresp}) in all possible ways into a hard and a soft 
subgraph. The hard subgraphs contribute to the Wilson coefficients 
and are usually calculated perturbatively, with the integration range 
of each internal momentum restricted\footnote{In practice, this 
restriction is often unnecessary (see below).} to be larger than the 
OPE scale $\mu \sim 0.5 \,{\rm GeV}$. The soft subgraphs correspond to 
hadron-channel independent condensates, renormalized at $\mu$. In the 
presence of an external electromagnetic field the OPE (up to 
eight-dimensional operators) involves the additional, Lorentz-covariant 
condensates
\begin{eqnarray}
\langle0|\bar{q} \sigma_{\mu\nu} q|0\rangle &=& \sqrt{4\pi\alpha} \chi 
F_{\mu\nu} \langle0|\bar{q} q|0\rangle ,  \label{magncond1} \\
g \langle0|\bar{q} G_{\mu\nu} q|0\rangle &=& \sqrt{4\pi\alpha} \kappa 
F_{\mu\nu} \langle0|\bar{q} q|0\rangle , \\
g \langle0|\bar{q} \gamma_5 \tilde{G}_{\mu\nu} q|0\rangle &=& 
\frac{i}{2} \sqrt{4\pi\alpha} \xi F_{\mu\nu} \langle0|\bar{q} q|0\rangle.
\end{eqnarray}
($G_{\mu\nu}=\frac12 \lambda_a G_{\mu\nu}^a$, $\tilde{G}_{\mu\nu} =
\frac12 \varepsilon_{\mu \nu \rho \sigma} G_{\rho \sigma}$ with 
$\varepsilon_{0123} = -1$.) The 
parameters $\chi, \kappa$, and $\xi$ play the role of generalized 
susceptibilities and quantify the vacuum response to weak 
electromagnetic fields. The magnetic susceptibility of the quark 
condensate, $\chi$, for example, originates from the induced spin 
alignment of quark-antiquark pairs in the vacuum. Note also that 
$\chi$ is associated with the lowest-dimensional induced condensate,
which enhances its role in the OPE and the corresponding sum rules. 

The OPE of $\Pi_{\mu\nu}(p)$ up to operators of dimension eight, with 
perturbatively calculated Wilson coefficients, has been obtained in 
Ref. \cite{IS}. An inherent assumption of this calculation - and of 
the QCD sum rule program in general - is that the short-distance 
physics associated with fields of wavelength smaller than $\mu^{-1}$ 
is predominantly perturbative. It is well known, however, that also 
strong nonperturbative fields of rather small size exist in the 
QCD vacuum. Instantons, i.e. the finite-action solutions of the 
classical, Euclidean Yang-Mills equation \cite{bel75}, are 
paradigmatic examples of such fields and have a crucial impact on 
the vacuum structure. 

The nonperturbative contributions from short-wavelength fluctuations 
of quarks and gluons around instantons are thus neglected 
in the standard treatment of the OPE. Their relative importance, 
and hence the justification for approximately disregarding them, 
depends both on the instanton size distribution in the vacuum and 
on the quantum numbers of the hadronic channel under consideration. 
Instantons of smaller (average) size $\bar{\rho}$ are accompanied 
by fluctuations of smaller wavelength, and those contribute more 
strongly to the Wilson coefficients. The hadron-channel dependence 
of the instanton contributions originates mainly from the chirality 
and spin-color coupling of the quark zero-modes in the instanton 
background. 

Instanton-induced effects are particularly large in the 
pseudoscalar-isovector and scalar-isoscalar channels, because there the 
quark zero-modes contribute with maximal strength. As a consequence, 
instanton contributions dominate already at short distances in 
the pseudoscalar sum rules, and are essential for their stability 
\cite{for95}. In the vector and axial-vector channels, on the other 
hand, zero-mode contributions are, to leading order in the instanton 
density, absent. 

The strength of instanton contributions to the nucleon channel lies  
about halfway in-between these extreme cases. In the chirally-odd 
amplitudes, it was found to be roughly of the same magnitude as 
that of the condensate contributions \cite{FB,for97}, since the 
spin-0 diquark operators in the interpolating fields (\ref{interpol}) 
couple strongly to instantons. The nucleon channel is therefore well 
suited for studying the interplay between instantons and other 
vacuum fields \cite{for97}. 

The calculation of the leading direct-instanton contributions to 
the background-field correlator (\ref{corr1}) proceeds essentially 
along the lines described in Refs. \cite{FB,for95,for97}, to which 
we refer for more details. Similar to the 
condensates, the bulk properties of the instanton size distribution 
are generated by long-distance vacuum dynamics and have thus to be 
taken as input for this calculation. As before \cite{FB,for95,for97}, 
we will use the standard values \cite{S1} $\bar{\rho} \simeq \frac{1}{3} 
\, {\rm fm}$ for the average instanton size and $\bar{R} \simeq 1 
{\rm fm}$ for the average separation between neighboring 
(anti)instantons. The results of instanton vacuum models 
\cite{sch98,SV} confirm these scales, while those from the lattice 
\cite{lat} are not yet fully consistent but lie in the same range 
(with maximal deviations of about 50 \%). 

Since the average instanton size is of the order of the inverse 
OPE scale, $\bar{\rho} < \mu^{-1} \simeq 0.4 \,{\rm fm}$, instanton 
corrections to the Wilson coefficients can be substantial. Moreover, 
since $\bar{\rho}^{-1} \gg \Lambda_{QCD}$, these corrections are 
essentially semiclassical, and at the relevant distances $x \lesssim 
0.2 \,{\rm fm} \ll \bar{R}$ multi-instanton correlations should be 
negligible. The instanton contributions to the OPE coefficients can 
therefore be calculated in semiclassical approximation, i.e. by evaluating 
the correlator (\ref{corr1}) in the background of the instanton and 
anti-instanton field and by then averaging the instanton parameters 
over their vacuum distributions. (The distribution of the instanton's 
position and color orientation is uniform, due to translational 
and gauge invariance). 

We treat the zero-mode sector of the quark propagator in the instanton 
field exactly and approximate the continuum modes, as before \cite{FB}, 
by plane waves. The recently found zero-mode dominance of the ground-state 
contributions to the pion and $\rho$-meson correlators on the lattice 
\cite{iva97} supports the validity of this approximation. The impact 
of the remaining vacuum fields (including other instantons) on the 
instanton contributions is accounted for in a mean-field sense 
\cite{shi80} and generates an effective mass $\bar{m} (\rho)= - 
\frac{2}{3} \pi^2\rho^2 \langle \bar{q} q \rangle$ for the quark zero 
modes. We further use the approximate instanton size distribution 
$n(\rho)=\bar{n}\delta(\rho-\bar{\rho})$ \cite{S1} and the self-consistency 
condition \cite{CDG} $\langle \bar{q} q \rangle = -2\int d\rho \, 
\frac{n(\rho)}{\bar{m} (\rho)} = -2 \, \frac{\bar{n}}{\bar{m} 
(\bar{\rho})}$ (which is numerically satisfied to good accuracy) 
to eliminate the $\bar{n}$ dependence from the resulting expressions.

Up to operators of dimension eight, we find the leading instanton 
contributions to the correlator (\ref{corr1}) to arise from just one 
type of graph, in which two of the quarks (emitted from 
the current (\ref{interpol}) at $x = 0$) propagate in zero-modes while 
the third interacts with the background field through the magnetized 
quark condensate. Graphs in which the background 
field couples directly to a hard quark in a zero-mode, vanish. The 
same holds for graphs in which the interaction with the photon causes 
a transition from zero-mode to continuum-mode 
propagation\footnote{Contributions of this type are essential in the 
pseudoscalar three-point 
correlator associated with the pion electromagnetic form factor 
\cite{for95}.}. The contribution from direct instantons is thus 
generated by the interplay between the rather localized quark zero modes 
and more slowly varying, nonperturbative vacuum fields. Effects of 
this type appear naturally in the OPE, whereas they are difficult to 
account for in, e.g., quark models \cite{for97}. 

Evaluating the corresponding graphs in the instanton and magnetic 
background fields as described above, and averaging over the instanton 
size distribution, we arrive at 
\begin{equation}
{\langle0|T \eta_p (x) \bar{\eta}_p (0)|0\rangle}_{F, inst} = \Pi_0^{inst} (x) 
-\frac{2^3 e_u}{3 \pi^4} \frac{\bar{\rho}^4}{\bar{m}^2} \langle \bar{q} 
\sigma_{\mu\nu} q \rangle_F \, \sigma_{\mu\nu} \int d^4 x_0 \frac{1}{
(r^2+\bar{\rho}^2)^3 (x_0^2+\bar{\rho}^2)^3} \label{instinx}
\end{equation}
($r = x - x_0$, where $x_0$ specifies the center of the instanton) for 
the proton. The corresponding neutron correlator is obtained by 
replacing $e_u$ with $e_d$. In order to put the instanton contribution 
(\ref{instinx}) to use in the sum rules, we also need its Fourier and 
Borel transform\footnote{For the Borel transform we follow the convention 
of Ref. \cite{I}.}. Again for the proton, it reads
\begin{equation}
\widehat{\Pi}_3 (M^2)=\frac{e_u}{128\pi^4}a\chi \bar{\rho}^2 
M^6 I(z^2). \label{incontr2}
\end{equation}
Here, $M$ denotes the Borel mass parameter. We have also used 
the standard definitions $z \equiv M \bar{\rho}$, $a \equiv-(2\pi)^2 
\langle \bar{q} q \rangle$, and abbreviated the integral
\begin{equation}
I(z^2)=\int_0^1\frac{d\alpha}{\alpha^2 (1-\alpha)^2}
e^{-\frac{z^2}{4\alpha(1-\alpha)}} = 4 e^{-\frac{z^2}{2}} \left[ 
K_0 \left( \frac{z^2}{2} \right) + K_1 \left( \frac{z^2}{2} \right) 
\right]. \label{instint}
\end{equation}
Note that (\ref{instint}) shows the typical exponential Borel-mass 
dependence of instanton contributions \cite{FB}. Together with the 
appearance of the new scale $\bar{\rho}$, this distinguishes them 
from the logarithms and power terms of the standard OPE.

An important qualitative property of the instanton contribution has 
been made explicit in Eq. (\ref{incontr2}): to leading order, direct 
instantons contribute almost exclusively\footnote{There is also a small 
direct--instanton contribution of similar structure to the amplitude 
$\Pi_2$. This term turns out to be too small to have an appreciable 
impact on the corresponding sum rules, however, and will not be 
discussed further.} to one invariant amplitude, $\Pi_3$, 
which is associated with the chirally-odd Dirac structure $\sigma_{\mu 
\nu}$. Exactly this amplitude was singled out in the previous sum rule 
analysis of Ref. \cite{IS}, for two reasons: first, one of its Wilson 
coefficients contains, in the ``pragmatic'' version of the OPE (see 
below), an infrared divergence. Secondly, the 
$\Pi_3$ sum rule failed to show a fiducial Borel region of stability 
\cite{ISP} (even after proper subtraction of the divergence), while 
the other two sum rules provide stable and accurate values for the 
nucleon magnetic moments even without direct-instanton corrections, 
and with a value of $\chi$ consistent with other, independent estimates 
\cite{BK}.

To understand the first point we note that, since in QCD perturbative 
contributions from soft loop momenta are normally small compared to 
the corresponding condensate contributions, it is standard procedure 
not to remove them in sum-rule calculations \cite{nov85}. This 
simplification - which goes under the name of ``pragmatic OPE'' - 
fails, however, if infrared divergences appear in diagrams associated 
with a Wilson coefficient. Such an infrared divergence was encountered 
in the OPE of $\Pi_3$, in a graph where a vacuum gluon field and the 
background photon interact with the same hard quark line. Hence the 
contribution from soft loop momenta has to be cut off explicitly, 
according to the rules of the exact OPE. A similar divergence was 
found before \cite{Sm} in the vector meson correlator when two soft 
gluon fields couple to the same quark line. 

The authors of \cite{IS} conjectured that the appearance of infrared 
singularities in $\Pi_3$ and the absence of a stability region in the 
associated sum rule might be related. This seems unlikely, however, in 
view of the later finding \cite{WPC} of similar infrared divergencies 
(which always occur when a quark line interacts with multiple 
soft gauge fields) in the other two amplitudes of (\ref{linresp}), which 
nevertheless lead to satisfactory sum rules. Direct instantons, on the 
other hand, contribute almost exclusively to $\Pi_3$, and their previous 
neglect could offer a more plausible explanation for the instability 
of the $\Pi_3$ sum rule. Support for this conjecture, which we are 
going to test quantitatively below, comes from two other chirally-odd nucleon 
sum rules (for the nucleon mass \cite{FB} and its isospin splitting 
\cite{for97}), where exactly such a selective stabilization due to 
instantons has been found. 

Our modified $\Pi_3$ sum rule is obtained by equating the Borel 
transform of the standard OPE from Ref. \cite{IS} and the instanton 
contributions (\ref{incontr2}) to the Borel-transformed double 
dispersion relation for the correlator (\ref{corr1}), with a 
spectral function parametrized in terms of 
the nucleon pole contribution (containing the magnetic moments) and 
a continuum based on local duality. Including the infrared-divergent 
term encountered in Ref. \cite{IS}, duly truncated at the OPE 
renormalization point, the new sum rule for the proton reads\footnote{We 
have corrected some errors which appeared in the expressions of Ref. 
\cite{IS}.}
\begin{eqnarray}
\nonumber
aM^2 \left\{ \left[ e_u-\frac{1}{6}e_d(1+4\kappa+2\xi) \right] E_1(M) 
\right.&+& \frac{1}{6}e_u\frac{m_0^2}{M^2} \left[\ln{\frac{M^2}{\mu^2}} 
- \gamma_{E M} \right] L^{-\frac{4}{9}} \nonumber \\
\left. +\frac{1}{6}e_d M^2\chi E_2(M) L^{-\frac{16}{27}}
- \frac{1}{8}e_u\chi\rho_c^2M^4I(z^2) L^{-\frac{16}{27}} \right\}
&=& \frac{1}{4}\tilde{\lambda}_N^2m
e^{-\frac{m^2}{M^2}} \left[\frac{\mu_p}{M^2}-\frac{\mu^a_p}{2m^2}
+A_p \right],
\label{sumrule}
\end{eqnarray}
where $m$ is the nucleon mass, $W$ the continuum threshold, and 
$\lambda_N$ the coupling of the current (\ref{interpol}) to the 
nucleon state, $\langle 0|\eta|N \rangle = \lambda_N u$. For the 
mixed quark condensate we use the standard parametrization $\langle 
\bar{q} \sigma_{\mu \nu} G^{\mu \nu} q \rangle = - m_0^2 \langle 
\bar{q} q \rangle$ with $m_0^2 = 0.8 \,{\rm GeV}^2$, and $\gamma_{E M}
\simeq 0.577$ is the Euler-Mascheroni constant. The additional 
parameters 
$A_{p,n}$ determine the strength of electromagnetically induced 
transitions between the nucleon and its excited states. The sum rule 
for the neutron is obtained from (\ref{sumrule}) by interchanging 
$e_u$ and $e_d$ and by replacing $\mu_p,\mu_p^a \rightarrow \mu_n$ 
and $A_p \rightarrow A_n$. We have also defined $\tilde{\lambda}_N^2=32 
\pi^4\lambda_N^2$, $L=\ln(M/\Lambda)/ \ln(\mu/\Lambda)$ ($\Lambda 
= 0.1 \,{\rm GeV}$), and transferred, using the standard expressions
\begin{eqnarray}
E_n(M)=1-e^{-\frac{W^2}{M^2}} \left[1+\sum_1^n\frac{1}{j !}\left(
\frac{W^2}{M^2}\right)^j \right], \label{cont}
\end{eqnarray}
the continuum contributions to the OPE-side of the sum rules. The 
appropriate form of these contributions has recently been clarified 
in Ref. \cite{iof95}.

In principle, several alternative options are available for the 
quantitative analysis of background-field sum rules. In practice, 
however, one is limited by the fact that their fiducial domain (i.e. 
the Borel-mass region in which the neglect of higher-order terms in 
the short-distance expansion is justified while the nucleon pole 
still dominates over the continuum) is generally not large enough to 
determine all the unknown parameters from a stable fit. The authors 
of Ref.\cite{IS} succeeded, however, in eliminating the susceptibilities 
and other constants by combining the two sum rules for $\Pi_1$ 
and $\Pi_2$ and their $M^2$-derivatives. The magnetic moments can then 
be fitted and are in good agreement with experiment, although taking 
derivatives of the sum rules generally reduces their reliability.  
Unfortunately, the above procedure also eliminates the continuum 
contributions, which makes it impossible to determine whether a 
fiducial stability domain exists. 

In any case, this procedure is ineffective for our new sum rules 
(\ref{sumrule}), as it cannot eliminate the additional $\chi$-dependence 
introduced by the direct-instanton contribution. Similarly, working 
with the ratio of background-field and mass sum rules, as done in Ref. 
\cite{chi86} for the $\Pi_1$ sum rules, offers no advantage in our case 
since already the leading terms in the OPE of the chirally-odd mass 
and background-field sum rules differ. Therefore, we resort to a direct 
minimization of the relative deviations between the two sides of Eq. 
(\ref{sumrule}). The coupling $\tilde{\lambda}^2 = 2.93 \,{\rm GeV}^6$ 
and the continuum threshold $W = 1.66 \,{\rm GeV}$ are, following the 
procedure of Ref. \cite{IS}, obtained by fitting the instanton-improved 
nucleon mass sum rule \cite{FB} to the experimental nucleon 
mass\footnote{We have included anomalous-dimension corrections in the 
mass sum rule of Ref. \cite{FB} and restored the full four-quark 
condensates, since we are dealing exclusively with Ioffe's interpolating 
field (\ref{interpol}).}. The values of the two susceptibilities $\kappa 
= -0.34 \pm 0.1$ and $\xi = -0.74 \pm 0.2$ were estimated in independent 
work by Kogan and Wyler \cite{KW}. This enables us to fit both sides of 
the sum rules (\ref{sumrule}) by varying $\chi$ and $A_p$ (or $A_n$, 
respectively) while keeping the magnetic moments fixed at their 
experimental values. 

The fits are performed in the fiducial Borel mass domain, which is 
bounded from below by requiring the highest-dimensional operators to 
contribute at most 10 \% to the OPE and from above by restricting the 
continuum contribution to maximally 50 \%. The resulting fiducial 
domains of both the proton and neutron sum rules, $0.8 \,{\rm GeV} 
\lesssim M \lesssim 1.15 \,{\rm GeV}$, are larger than those of the 
sum rules based on $\Pi_1$ and $\Pi_2$. 

The fit results are shown in Fig. \ref{fig1}, where for both the proton 
and the neutron sum rules the direct-instanton contributions, 
the remaining OPE including the continuum contributions, their sum 
(which makes up the left-hand side of Eq. (\ref{sumrule})), and the 
right-hand sides are plotted. The fit quality of both the proton and 
neutron sum rules is excellent. As previously in the instanton-improved, 
chirally-odd nucleon mass sum rule \cite{FB}, the theoretical side of 
the sum rules, including the instanton-induced part, is almost 
indistinguishable from the phenomenological side. Moreover, Fig. 
\ref{fig1} shows that the direct-instanton contributions can reach 
about half the magnitude of the remaining terms in the OPE, which 
makes it evident why their previous neglect had a detrimental 
impact on the stability properties. 

An alternative way of evaluating the optimized sum rules consists in 
solving them for $\mu_N$ and plotting the result as a function of the 
Borel mass, as shown in Fig. \ref{fig2}. The resulting functions 
$\mu_{p,n} (M)$ specify the value of the magnetic moment which is 
required to make both sides of the sum rule (\ref{sumrule}) match 
exactly at each value of the Borel mass. The instanton-corrected sum 
rules render $\mu(M)$ 
practically $M$-independent, thereby again indicating an almost 
perfect fit for both proton and neutron. 

This fit predicts $\chi\simeq -4.96 \, {\rm GeV}^{-2}$ for the proton 
and $\chi\simeq -4.73 \, {\rm GeV}^{-2}$ for the neutron sum rule. These 
values correspond to the OPE scale $\mu = 0.5 \,{\rm GeV}$ adopted for 
our sum rules and lie inside the range obtained from other estimates 
\cite{IS,BK,WPC}. They are somewhat smaller in magnitude than the value 
$\chi \simeq - 5.7 \, {\rm GeV}^{-2}$ found in the two- and three-pole 
models of Ref. \cite{BK}. (Our predicted values for the excited-state 
transition parameters are $A_p \simeq 0.28 {\rm GeV}^2$ and $A_n 
\simeq -0.27 {\rm GeV}^{-2}$.)

In conclusion, we have recovered a third reliable sum rule for the 
nucleon magnetic moments. In contrast to the other two, it receives 
previously neglected direct-instanton contributions which arise 
from the interplay with long-wavelength vacuum fields. 
Our new sum rule is built on the chirally-odd amplitude $\Pi_3$ 
of the nucleon correlator in an electromagnetic background field and 
found to be at least as stable as the other two, although it had 
previously been regarded as flawed. The new sum rule adds 
to the predictive power of the background-field sum rules and 
strengthens their mutual consistency. 

Furthermore, our results reinforce a systematic pattern which emerged 
from previous studies of direct-instanton effects both in the nucleon 
and pion channels: those sum rules which worked satisfactorily 
without instanton corrections receive little or no direct instanton 
contributions, and previously less reliable or completely unstable sum 
rules are stabilized by large instanton contributions. This pattern 
points not only towards the importance of direct instantons in 
particular sum rules, but also supports the adequacy of their 
semiclassical implementation into the OPE. Our results show that 
these conclusions continue to hold in the presence of a ``magnetized'' 
vacuum. 

M.A. and M.K.B. acknowledge support from the U.S. Dept. of Energy under 
grant number DE-FG02-93ER-40762. H.F. acknowledges support from Deutsche 
Forschungsgemeinschaft under habilitation grant Fo 156/2-1.

\begin{figure}
\caption{The OPE (dashed line) and direct instanton (dotted line) 
contributions to the new $\sigma_{\mu\nu}$ sum rules for the proton 
(positive range) and neutron. Their sum (dot-dashed line) is compared 
to the RHS (solid line).}
\label{fig1}
\end{figure}
\begin{figure}
\caption{The Borel mass dependence of the magnetic moments of the 
proton (upper) and neutron calculated from the optimal fit of LHS and 
RHS.}
\label{fig2}
\end{figure} 
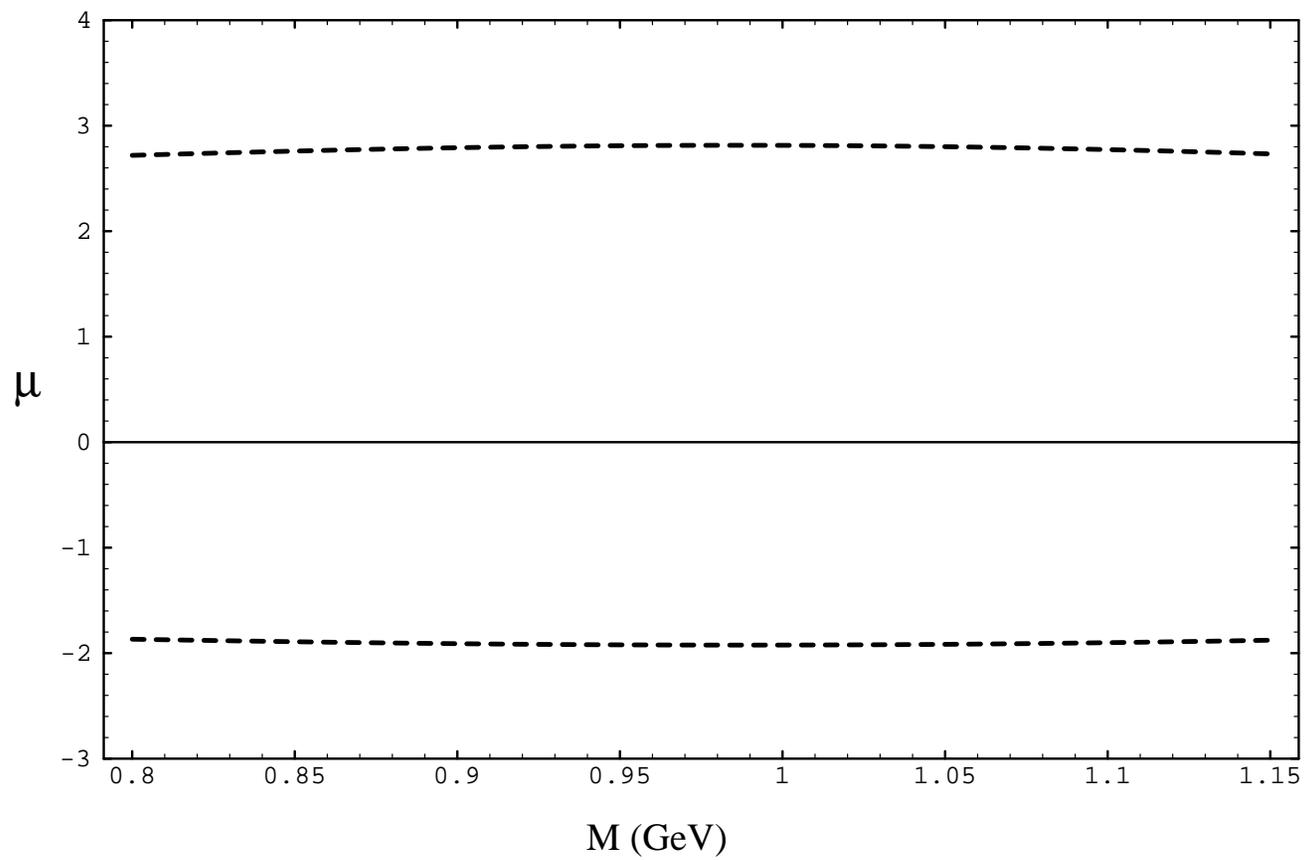
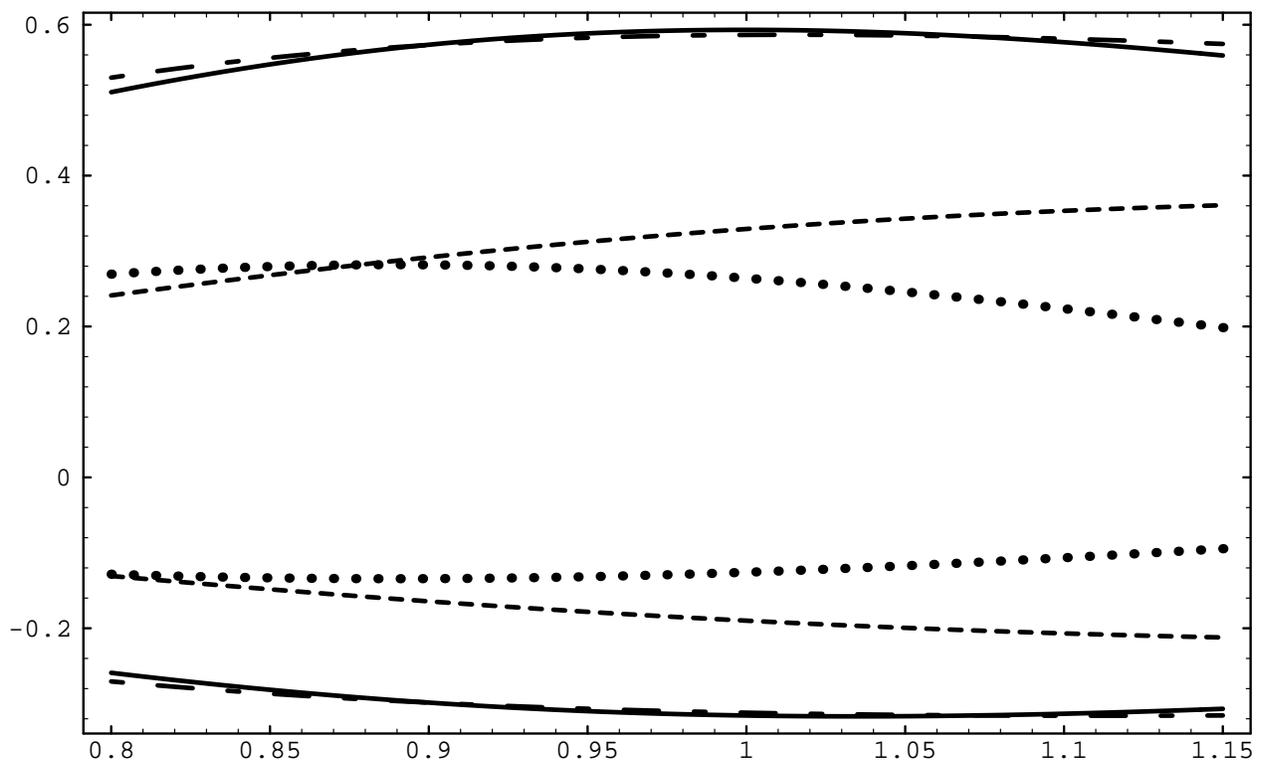
\end{document}